\documentclass[11pt]{article}
\usepackage{graphicx}
\def \d {{\rm d}}

\begin{document}

\title{Accelerating black holes in  anti-de~Sitter universe
\thanks{Dedicated to my academic teacher Prof. J.~Bi\v{c}\'ak on the
occasion of his $60^{th}$ birthday.}}

\author{J. Podolsk\'y
\\
\\ Institute of Theoretical Physics, Charles University in Prague,\\
V Hole\v{s}ovi\v{c}k\'ach 2, 180\ 00 Prague 8, Czech Republic.\\
{\small E--mail: {\tt podolsky@mbox.troja.mff.cuni.cz}} }

\date{\today}
\maketitle

\begin{abstract}
A physical interpretation of the $C$-metric with a negative cosmological
constant $\Lambda$ is suggested. Using a convenient  coordinate system
it is demonstrated that this class of exact solutions of Einstein's
equations describes uniformly accelerating (possibly charged) black
holes in anti-de~Sitter universe. Main differences from the
analogous de~Sitter case are emphasised.
\end{abstract}


\section{Introduction}

The well-known $C$-metric \cite{EhKu62}, \cite{KSMH} is an important explicit
representative of a class of spacetimes with boost-rotation symmetry \cite{BicSch89}.
It was shown already in 1970 by Kinnersley and Walker \cite{KinWal70} that this
describe a pair of uniformly accelerating black holes of masses~$m$ and possible
charges~$e$ in Minkowski background. The acceleration $A$ is caused by conical
singularities which may be interpreted either as a strut between the two black holes
or  two semi-infinite strings connecting them to infinity. Geometrical and asymptotic
properties were subsequently studied in \cite{FarZim80}, \cite{AsDr81}. Bonnor \cite{Bon83}
found a transformation of the $C$-metric into the  boost-rotational canonical
form. However, the explicit metric functions are
somewhat complicated and  depend on specific ranges of the initial ``static''
coordinates  \cite{Bon83}-\cite{PraPra00}. Recently the  limit of
unbounded acceleration  $A\to\infty$ was investigated \cite{[B12]}.
It has been demonstrated that such limit of the $C$-metric is identical to the
solution which represents a spherical  impulsive gravitational
wave generated by a snapping string (or an expanding strut)
\cite{GlePul89}-\cite{NutAli}.

Interestingly, there exists a more general class of the $C$-metric solutions
presented  by Plebanski and  Demianski \cite{PleDem76}
which admits a non-vanishing value of a cosmological constant
$\Lambda\>$ (plus possibly a rotational parameter, see also
\cite{Carter68}-\cite{LeOl01}). These solutions have already been used
for investigation of the pair creation of black holes in  de~Sitter
\cite{ManRos95} and anti-de~Sitter backgrounds \cite{Mann97}.
It is also natural to expect that in the limit \  $A\to\infty$ \
this generalised $C$-metric generates a specific spherical
impulsive gravitational wave in the (anti-)de~Sitter universe which
may be interesting in the context of quantum and string theories.
However, physical and global properties of the spacetimes
\cite{PleDem76} with \ $\Lambda\ne0$ \ have not yet been investigated
thoroughly even at the classical level. Only the first steps have
 been done. Using a suitable coordinate system adapted to
the motion of uniformly accelerating particles in de~Sitter space
the physical meaning of the parameters appearing in solutions \cite{PleDem76}
could be determined  \cite{[B13]}. On the other hand, the most recent paper by
Bi\v{c}\'ak and Krtou\v{s} \cite{BicKr} on accelerated test sources
in de Sitter spacetime may be of a great help for understanding the
global structure of these solutions.

It is the purpose of this short contribution to provide a basic physical
interpretation of the above family of exact solutions with a {\it negative}
value of the cosmological constant $\Lambda$ as  uniformly accelerating black
holes in anti-de~Sitter universe. Although this is a simple extension of
our previous results presented in \cite{[B13]}, we shall
demonstrate that there are some interesting new features and
fundamental differences from the analogous de~Sitter case.

\section{Uniformly accelerating observers in anti-de Sitter space}

Anti-de~Sitter universe is the (maximally symmetric)  spacetime of constant negative curvature
\cite{HE}. It can be understood as a 4-dimensional hyperboloid
   \begin{eqnarray}
  -Z_0^2+Z_1^2+Z_{23}^2-Z_4^2=-a^2\>,
  \label{fourhyp}
   \end{eqnarray}
where $a=\sqrt{-3/\Lambda}$,
in a 5-dimensional Minkowski space
\begin{equation}
\d s^2=-\d Z_0^2+\d Z_1^2+\d Z_{23}^2+Z_{23}^2\,\d \Phi^2-\d Z_4^2\ ,\label{ds0}
\end{equation}
as shown in Fig.~1.
We have introduced here for later convenience the coordinates
$Z_{23}\in[0,\infty)$ and $\Phi\in[0,2\pi)$ by
\ $Z_2=Z_{23}\cos\Phi\>$\  and\  $Z_3=Z_{23}\sin\Phi$, instead of the standard
Cartesian coordinates $Z_2$, $Z_3$. With the parameterisation
  \begin{eqnarray}
  &&Z_0 =a\>{\sin(T/a)\over\cos\chi}\>,\qquad\
  Z_4   =a\>{\cos(T/a)\over\cos\chi} \>, \nonumber\\
  &&Z_1 =a\>\tan\chi\,\cos\Theta \>,\quad
  Z_{23}=a\>\tan\chi\,\sin\Theta \>, \label{global}
  \end{eqnarray}
where $T\in(-\infty,+\infty)$, $\chi\in[0,{\pi\over2})$,
$\Theta\in[0,\pi]$, we obtain the well-known metric in  global coordinates
 \begin{equation}
 \d s^2 ={a^2\over\cos^2\chi} \left[-\d\, (T/a)^2+ \d\chi^2
  +\sin^2\chi\,(\d \Theta^2+\sin^2\Theta\,\d\Phi^2)\right] \>.
  \label{conffmetr}
 \end{equation}
This immediately gives the familiar Penrose diagram for the anti-de~Sitter
space (see Fig.~1) in which the boundary $\chi={\pi\over2}$ represents the conformal
infinity ${\cal J}$ for null and spacelike geodesics \cite{HE}.

Let us consider the timelike worldlines in (\ref{conffmetr}) of the form
\begin{equation}
T=\tau\,\cos\chi_0\>,\ \  \chi=\chi_0\>,\ \  \Theta=\Theta_0\>,\ \ \Phi=\Phi_0\>,
\label{worldlines}
\end{equation}
where $\tau$ is the proper time, and $\chi_0$, $\Theta_0$, $\Phi_0$ are constants.
The corresponding 4-velocity is \ $u^\mu=(\cos\chi_0,0,0,0)$, \ and the 4-acceleration is
\ $\dot u^\mu\equiv u^\mu_{\ ;\nu}u^\nu=(0,a^{-2}\sin\chi_0\cos\chi_0,0,0)$, with the
constant modulus
  \begin{equation}
   |A|\equiv|\,\dot u^\mu|={\sin\chi_0\over a} \ .
  \label{A}
  \end{equation}
Moreover, due to the relation  \ $\,u_\mu \dot u^\mu=0$, this constant value of $|A|$
is identical to the modulus of the 3-acceleration measured in the natural local
frame orthogonal to the observer's 4-velocity $u^\mu$. The family of worldlines
(\ref{worldlines}) thus represents the motion of {\it uniformly accelerating observers
in the anti-de~Sitter universe}. Their trajectories
are given  by simple vertical lines in the Penrose diagram shown in Fig.~1.
 For \ $\chi_0=0$ \
the worldline is a geodesic with zero acceleration. With growing values of
$\chi_0$ the acceleration of the corresponding observers given by
(\ref{worldlines}) uniformly grows.
Interestingly, there exists the {\it maximal possible value} $A_{max}$,
  \begin{equation}
|A|\le A_{max}\equiv \sqrt{-{\textstyle{1\over3}}\Lambda}\ , \label{Amax}
  \end{equation}
of uniform acceleration of the privileged observers (\ref{worldlines}) in the
anti-de~Sitter space which is reached in the limit $\chi_0={\pi\over2}$, i.e.
for those moving ``along the conformal infinity''.
This is completely different from the situation in the de~Sitter universe,
in which case the  acceleration $A$ of analogous  observers may be unbounded,
see e.g. \cite{ManRos95}, \cite{[B13]}, \cite{BicKr}. Note that such infinite
acceleration corresponds to (null) observers in the de~Siter universe which move
along the cosmological horizon. This is absent in the anti-de~Sitter space.

In the 5-dimensional formalism the uniformly accelerated trajectories (\ref{worldlines})
are also privileged. It immediately follows from (\ref{global}) that these are  given by
constant values of $Z_1$ and $Z_{23}$. Therefore, the trajectories are just simple closed
timelike loops of the radius $\>a/\cos\chi_0\>$ around the anti-de~Sitter hyperboloid
 indicated in Fig.~1.

We may now introduce a new coordinate system which is well adapted to description
of uniformly accelerating  test point sources in the anti-de~Sitter
space. This  analogue of the accelerated coordinates for the de~Sitter
spacetime \cite{[B13]} is given by the following parameterisation of
the hyperboloid (\ref{fourhyp}),
   \begin{eqnarray}
  &&Z_0={\sqrt{a^2+r^2}\,\sin(T/a) \over
      \sqrt{1-a^2A^2} + A\,r\cos\theta} \ , \qquad\quad
  Z_4={\sqrt{a^2+r^2}\,\cos(T/a) \over
      \sqrt{1-a^2A^2} + A\,r\cos\theta}\ , \nonumber\\
  &&Z_1={\sqrt{1-a^2A^2}\,\,r\cos\theta-a^2A\over
      \sqrt{1-a^2A^2} + A\,r\cos\theta} \ , \ \>\quad
  Z_{23}={r\sin\theta \over
      \sqrt{1-a^2A^2} + A\,r\cos\theta}\ , \label{acceler}
  \end{eqnarray}
where \ $r\in[0,\infty)$, \ $T\in(-\infty,\infty)$, and  \
$\theta\in[0,\pi]$.  In these coordinates, the anti-de~Sitter space
takes the form
  \begin{equation}
\d s^2 = {\ 1\over[\sqrt{1-a^2A^2} + A\,r\cos\theta\>]^2} \Bigg\{
 -{ \left(1+{r^2/ a^2}\right) }\,\d T^2
 +{\d r^2\over { 1+{r^2/a^2}}}
+r^2(\d \theta^2+\sin^2\theta\,\d\Phi^2) \> \Bigg\} \>.
  \label{accmetr}
  \end{equation}
In the case $A=0$ this  static metric reduces to the standard form
(\ref{conffmetr})  if we perform obvious transformations
$\,r=a\tan\chi\,$, and $\,\theta=\Theta$.  Moreover, the metric (\ref{accmetr})
is conformal to this for a
general value of the acceleration $A<A_{max}$.  The explicit transformation
between (\ref{conffmetr}) and (\ref{accmetr}) is
obtained by comparing (\ref{global}) with (\ref{acceler}),
  \begin{eqnarray}
\cos\chi &=&  {\sqrt{1-a^2A^2} + A\,r\cos\theta \over
     \sqrt{1+r^2/a^2}}       \ , \nonumber\\
a\tan\chi\sin\Theta &=& {r\sin\theta \over
      \sqrt{1-a^2A^2} + A\,r\cos\theta}\ , \label{rtheta}\\
a\tan\chi\cos\Theta &=& {\sqrt{1-a^2A^2}\,\,r\cos\theta-a^2A\over
      \sqrt{1-a^2A^2} + A\,r\cos\theta} \ ,  \nonumber
  \end{eqnarray}
which relate  $\chi,\Theta$ to $r,\theta$ (the coordinates $T$ and $\Phi$ have
the same meaning in both the metrics). It is obvious from the first relation
in (\ref{rtheta}) that the origin \ $r=0$ \ of the coordinates
(\ref{accmetr}) corresponds to \ $\cos\chi_0=\sqrt{1-a^2A^2}$, i.e.
\ $\sin\chi_0=a|A|\,$. \ From (\ref{A}) we immediately
conclude that the parameter $A$ in the metric (\ref{accmetr}) is exactly
the value of the acceleration of the corresponding observer. Therefore,
{\it the origin \ $r=0$ \ of the coordinates in} (\ref{accmetr}) {\it is
accelerating in anti-de~Sitter universe with uniform acceleration $A$}.
It also follows from (\ref{rtheta}) that the coordinate singularity
\ $r=0$ \ is located at \ $\Theta_0=0$ when \ $A<0\,$, whereas  it is
located at \ $\Theta_0=\pi\,$ when \ $A>0\,$.
Considering (\ref{acceler}) we finally  observe that the trajectory of
uniformly accelerating  origin \  $r=0$ \  corresponds to a uniform
motion of a single point around the anti-de~Sitter hyperboloid. This
closed trajectory is a circle of radius \  $a/\sqrt{1-a^2A^2}$, \ which is given
by an intersection of the hyperboloid (\ref{fourhyp}) with the
plane having the constant values  \ $Z_1=-a^2A/\sqrt{1-a^2A^2}$\
and \  $Z_{23}=0\,$, see Fig.~1.

\section{Physical interpretation of the $C$-metric with $\Lambda<0$}

The Plebanski--Demianski solution for the $C$-metric with
a cosmological constant  can be written in the form \cite{PleDem76}
  \begin{equation}
  \d s^2 = {1\over(\,p+q\,)^2} \left({\d p^2\over{\cal P}}
+{\d q^2\over{\cal Q}} + {\cal P}\>\d\sigma^2  -{\cal Q}\>\d\tau^2 \right)\ ,
  \label{Cmetric}
  \end{equation}
  where
  \begin{eqnarray}
  &&{\cal P}(p) = A^2 -p^2 + 2m\,p^3 -e^2p^4\ ,   \label{PQ}\\
  &&{\cal Q}(q) = {1\over a^2}-A^2 +q^2 + 2m\,q^3 +e^2q^4\ . \nonumber
  \end{eqnarray}
Note that we do not consider the possible rotation here. Also, we
have used the coordinate freedom to remove the linear terms  in (\ref{PQ}),
and to set the coefficients of the quadratic terms  to unity.
In order to maintain the spacetime signature, it is necessary
that ${\cal P}>0$ which places a restriction on the range of $p$.
However, there is no restriction on the sign of ${\cal Q}$ which may
describe both static and non-static regions, separated by horizons localised
on ${\cal Q}=0$. Using the standard ``Descartes sign rule'' we
 obtain for\ $\>m>0\>$ and  $\,|A|<A_{max}\,$ that the
polynomial ${\cal P}(p)$  given by (\ref{PQ}) has at most {\it four}
real roots $p_i$ (three positive and one negative). On the other hand,
the polynomial ${\cal Q}(q)$ has at most {\it two} real (necessarily
negative) roots $q_i$. There are thus four possible static spacetime regions
for all permitted values of $p$ and $q$, as illustrated in Fig.~2 by the
shaded areas. These are bounded by horizons $q_i$ and ``coordinate''
singularities $p_i$. In the following we shall concentrate on the
spacetime for which \ $p_1<p<p_2$ \ and \ $p+q<0$\  (note that  \ $p+q=0$ \
corresponds to the spacetime boundary on which the conformal factor in
(\ref{Cmetric}) becomes  unbounded, whereas $q=-\infty$ represents
a curvature singularity).  This covers one non-static plus two
static regions separated by inner and outer black hole
horizons, $q=q_1$ and $q=q_2$, respectively. Again, this is different
from the  de~Sitter case for which {\it both} the polynomials
 ${\cal P}$ and  ${\cal Q}$ may have up to four real roots so that an additional
cosmological horizon (on $q=q_3<0$) is present.

Let us perform the coordinate transformation of
the metric (\ref{Cmetric}) given by
  \begin{eqnarray}
  && T=\sqrt{1-a^2A^2}\>\tau \ , \qquad
    r=-{\sqrt{1-a^2A^2}\over q} \ , \nonumber \\
  && \theta=\int_{\zeta_1}{\d \zeta\over\sqrt{
1-\zeta^2+2mA\zeta^3-e^2A^2\zeta^4}}  \ , \qquad
  \Phi={A\over c}\,\sigma\ ,  \label{transf}
  \end{eqnarray}
 where  \ $\zeta\equiv p/A\,$, and $c$ is a constant. We obtain
  \begin{equation}
\d s^2 = {\ 1\over[\sqrt{1-a^2A^2} - A\,r\,\zeta(\theta)\,]^2}\,
 \Bigg\{ -   F(r)\,\d T^2 + {\d r^2\over F(r)} +r^2\Big(\,\d \theta^2
+G^2(\theta)\,c^2\d\Phi^2 \Big) \Bigg\}\  ,  \label{newmetric}
  \end{equation}
in which
  \begin{eqnarray}
 F(r)&=&1 +{r^2\over a^2} -\sqrt{1-a^2A^2}\,{2m\over r}
+(1-a^2A^2)\,{e^2\over r^2} \ , \label{FG}\\
 G^2(\theta) &=&1 -\zeta^2(\theta) +2mA\,\zeta^3(\theta)
-e^2A^2\,\zeta^4(\theta) \ ,\nonumber
  \end{eqnarray}
  and $\zeta(\theta)$ is the inverse function of $\theta(\zeta)$ given by
the integral in (\ref{transf}). It may be seen that, either when \ $A=0$ \
or when both \ $m=0$ \ and \ $e=0$, \ we obtain \
$\zeta(\theta)=-\cos\theta$ \ so that \ $G(\theta)=\sin\theta$. \ Otherwise
these can be expressed in terms of Jacobian elliptic functions
\cite{[B13]}.

It is obvious that for $A=0$, $c=1$ the metric (\ref{newmetric})
exactly reduces to the familiar form of the Reissner--Nordstr{\o}m--anti-de~Sitter
black hole solution in which the parameters $m$ and $e$ have the usual
interpretation and the curvature singularity is located at $r=0$.
Moreover, when $A\ne0$, $c=1$ and $m=0=e$, the line element (\ref{newmetric}) is
{\it identical} to (\ref{accmetr}) in which $r=0$ corresponds to a uniformly
accelerating single point in the anti-de~Sitter background. When $m$ and $e$
are small, the solution (\ref{newmetric}) can naturally be regarded as a
perturbation of (\ref{accmetr}). The metric (\ref{newmetric}) can thus be
interpreted as describing {\it a charged black hole which is uniformly
accelerating in anti-de~Sitter universe}.

The spacetimes given by the metric (\ref{newmetric}) explicitly possess the boost and
rotation symmetries corresponding to  the Killing vectors $\partial_T$ and $\partial_\Phi$.
There are Killing horizons where the norm of the Killing vector $\partial_T$
vanishes. These occur on $F=0$ and separate static and non-static regions of the
spacetimes. Generally,  for $r>0$, $m>0$ there are at most two real roots of $F(r)$
which correspond to the familiar inner and outer black hole horizons in the
Reissner--Nordstr{\o}m--anti-de~Sitter spacetimes, see e.g. \cite{Lake}-\cite{BiPo97}.
Their specific geometrical properties depend not only on the parameters $m$, $e$, and
$\Lambda$, but also on the acceleration $A$. Contrary to the de~Sitter background
there is no cosmological horizon. Cases representing uniformly accelerating
extreme black holes or  naked singularities in anti-de~Sitter universe are also
described by (\ref{newmetric}) for specific ranges of the parameters
such that the two possible roots of $F$ are repeated or absent.

For $A\not=0$ and a  general choice of the parameters $m$, $e$, $\Lambda$, the complete
global structure of the spacetime is  complicated. Of course, for small $m$ and $e$,
the solutions (\ref{newmetric}) can be considered as perturbations of anti-de~Sitter
universe illustrated in Fig.~1. However, these pictures now are only
schematic   since \ $r=0$ \ is not a single
``test'' point but a curvature singularity, and the black hole horizons also occur.
For vanishing acceleration $A$, the conformal diagrams of the corresponding
(spherically symmetric) spacetimes are already known \cite{Lake}-\cite{BiPo97}.
Interestingly, these diagrams also describe the global structure of the complete
 spacetime (\ref{newmetric}) {\it in the ``equatorial'' plane $\zeta(\theta)=0$} orthogonal to the
 direction of acceleration,  even in the case when $A\ne0$.
On this plane, $r=\infty$ corresponds to the anti-de~Sitter conformal infinity ${\cal J}$
which is represented by the  boundary $\chi={\pi\over2}$ in Fig.~1.

We finally clarify the character of the singularities at \
$\zeta_1=p_1/A$ \ and \ $\zeta_2=p_2/A$ \  which are the roots
of \ $G(\zeta)=0$.
It follows from the integral in (\ref{transf}) that \ $\zeta_1$ \
corresponds to \ $\theta=0$, \ whereas \ $\zeta_2$ \ to \ $\theta=2K$,
in which  $K$ is the ``quarter period'', i.e. the complete elliptic integral
of the first kind related to (\ref{transf}). (When either \ $A=0$ \ or
\ $m=0=e\>$ we  obtain $K={\pi\over2}$.) Therefore, the range of the
``angular'' coordinates in the metric (\ref{newmetric}) is
\ $\theta\in[0,2K]$, \ $\Phi\in[0,2\pi)$.

As in the de~Sitter case \cite{[B13]} we may  compare the circumference
of a small circle around the pole $\theta=0$ with its radius
(for any fixed value of $r$ and $T$). We find that in general there is a deficit angle
  \begin{equation}
  \delta_1
= 2\pi \left[\, 1-\lim_{\theta\to0}\,{c\,G(\theta)\over\theta}\,\right]
= 2\pi \left[\, 1-c\,G'(0) \right]\ , \label{delta1}
  \end{equation}
where \  $G'(0)=-\zeta_1+3mA\,\zeta_1^2-2e^2A^2\,\zeta_1^3\> $,
which is finite and independent of $r$ and $T$. Thus, the
singularity at \ $\zeta_1$ \ represents a {\it cosmic string of constant
tension} along the semi-axis  \ $\theta=0$. \ Of course, for any value of
the physical parameters $m$, $e$, $A$ and $\Lambda$, this can always be made
regular by putting \ $c=1/G'(0)$. \ In particular, if  \ $A=0$
(the case of spherically symmetric Reissner--Nordstr{\o}m--anti-de~Sitter spacetime),
\ or if $m$ and $e$ are both zero (the anti-de~Sitter universe in accelerating coordinates),
the semi-axis  \ $\theta=0\>$ is regular when \ $c=1\>$ because
\ $G=\sin\theta\>$ in these cases.

The deficit angle of the second cosmic string which extends in the opposite direction
 \ $\theta=2K\>$ is
  \begin{equation}
  \delta_2
= 2\pi \left[\, 1-\lim_{\theta\to2K}\,{c\,G(\theta)\over2K-\theta} \,\right]
= 2\pi \left[\, 1+c\,G'(2K) \,\right]\ , \label{delta2}
  \end{equation}
where \ $G'(2K)=-\zeta_2 +3mA\,\zeta_2^2 -2e^2A^2\,\zeta_2^3 \>$.
Again, the string along the semi-axis  \ $\theta=2K\>$ can be removed by
setting \ $c=-1/G'(2K)$. \ However, it is not possible in general to remove
the strings in both directions simultaneously unless \ $G'(0)=-G'(2K)\>$.
This condition can only be satisfied for \ $m=0=e\>\,$ or \ $A=0\>$
(in which case \ $\zeta_2=1=-\zeta_1\>$)  with \ $c=1$.
Thus, uniformly accelerating black hole in anti-de~Sitter universe
must necessarily be connected to at least one cosmic string
localised at \ $\theta=0\>$ and/or \ $\theta=2K\,$, which may be
 considered to ``cause'' the acceleration (see also \cite{Mann97}).
This is in accordance with  the same result obtained in Minkowski
\cite{KinWal70} or de~Sitter \cite{ManRos95}, \cite{[B13]} space.
There are no (non-trivial)
self-accelerating black holes in the backgrounds of constant curvature.

However, there are fundamental geometrical differences between the cases with
positive and negative value of the cosmological constant
$\Lambda$. In the de~Sitter case, there are {\it two} uniformly
accelerating black holes connected each other by   finite
string(s). In the presence of both strings these make a complete
{\it circular loop} around the whole  closed de~Sitter universe. In the
anti-de~Sitter background, there is just a {\it single} uniformly
accelerating black hole attached to  {\it semi-infinite} open
string(s). These have a ``hyperboloidal'' character and connect the black hole
to the anti-de~Sitter conformal infinity ${\cal J}$. Indeed, if we
treat the solution (\ref{newmetric}) for sufficiently small \ $m\>$ and
\  $e\>$ as a perturbation of the anti-de~Sitter hyperboloid
(\ref{fourhyp}), the first string is localised along
\ $\theta=0\>$ whereas the second along \ $\theta=\pi\>$.
Considering (\ref{acceler}) we observe that both these cases correspond
to \ $Z_{23}=0\>$. From (\ref{fourhyp}) it follows that the geometry of
such strings in anti-de~Sitter universe is  given by a hyperboloidal
2-surface \ $ (Z_0^2+Z_4^2)-Z_1^2=a^2\,$. The position of the strings
at a given time  is  indicated in Fig.~1  by the wavy line. The first
semi-infinite string localised on
\ $\theta=0\>$ connects the uniformly accelerated source \ $r=0\>$ to the
infinity ${\cal J}$ on the ``left'', \ $Z_1>-a^2A/\sqrt{1-a^2A^2}\>$,
while the second string on \ $\theta=\pi\>$ extends from the source in
the opposite direction to the ``right'', \ $Z_1<-a^2A/\sqrt{1-a^2A^2}\>$.

\section*{Acknowledgements}

This work was supported  by the grant GACR-202/99/0261 of the Czech Republic
and GAUK~141/2000 of Charles University in Prague.


\newpage

\vskip20mm

\begin{figure}
\centering
\caption
{The anti-de~Sitter space represented as a hyperboloid for the section $Z_{23}=0$
(left), and  the corresponding Penrose conformal diagram  (right). The trajectories
of observers with uniform acceleration $A$ given by $\chi=\chi_0$
(\,i.e., $\,Z_1=-a^2A\sqrt{1-a^2A^2}\,$, or $\,r=0$ ) are indicated. The
possible two semi-infinite cosmic strings localised on $Z_{23}=0$
extend from the observer in opposite directions $\theta=0$ and
$\theta=\pi$.
}
 \label{Fig.1}
 \end{figure}

\vskip20mm

 \begin{figure}
\centering
\caption
{In the full ranges of $p$ and $q$, physical spacetimes only occur when
$p\in(p_1,p_2)$, or $p\in(p_3,p_4)$, where $p_i$ are possible real roots of the
polynomial ${\cal P}\,$. There exist at most two real (necessarily
negative) roots $q_1$ and $q_2$ of the polynomial ${\cal Q}\,$.
Generically, there are thus four static spacetime regions which are indicated
by the shaded areas. We concentrate here on the spacetime spanned
by \ $p_1<p<p_2$ \ and \ $q<-p$ between the curvature singularity  $q=-\infty$
and the conformal infinity \ $p+q=0$. \ This contains one non-static and two static
regions which are separated by  inner ($q=q_1$) and outer ($q=q_2$) black hole
horizons.
}
 \label{Fig.2}
 \end{figure}

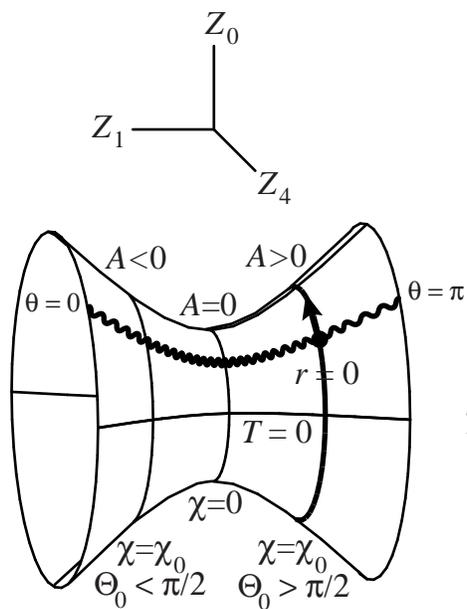
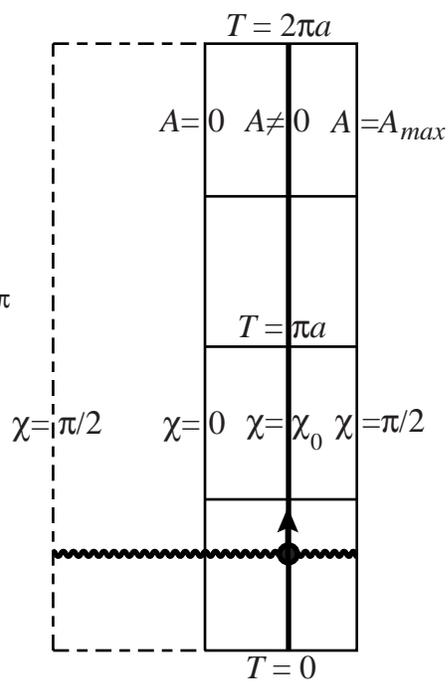
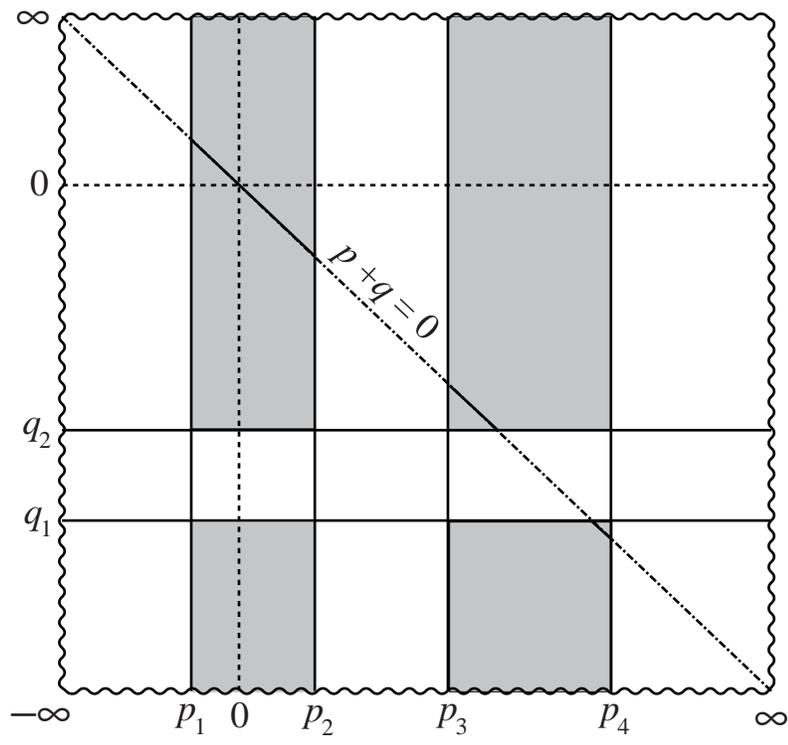
\end{document}